# Passivation mechanisms and pre-oxidation effects on model surfaces of FeCrNi austenitic stainless steel

*Li Ma,* [*] *Eirini-Maria Pascalidou,*[†] *Frédéric Wiame, Sandrine Zanna, Vincent Maurice,*[‡] *Philippe Marcus*[§]

PSL Research University, CNRS - Chimie ParisTech, Institut de Recherche de Chimie Paris (IRCP), Physical Chemistry of Surfaces Group, 11 rue Pierre et Marie Curie, 75005 Paris, France.

## Abstract

Passivation mechanisms were investigated on (100)-oriented Fe-18Cr-13Ni surfaces with direct transfer between surface preparation and analysis by X-ray photoelectron spectroscopy and scanning tunneling microscopy and electrochemical characterization. Starting from oxide-free surfaces, pre-oxidation at saturation under ultra-low pressure (ULP) oxygen markedly promotes the oxide film Cr(III) enrichment and hinders/delays subsequent iron oxidation in water-containing environment. Exposure to sulfuric acid at open circuit potential causes preferential dissolution of oxidized iron species. Anodic passivation forces oxide film re-growth, Cr(III) dehydroxylation and further enrichment. ULP pre-oxidation promotes Cr(III) hydroxide formation at open circuit potential, compactness of the nanogranular oxide film and corrosion protection.

---

[*] Present addresses: NO.21 Building C of the Hengqin Creative Valley, 1889 East Huandao Road, Hengqin New Area, Zhuhai Guangdong, China; Department of Materials Science, Fudan University, Shanghai, 200433, China

[†] Present address: Department of Chemistry-Ångström, Uppsala University, Uppsala, SE-751 21, Sweden

[‡] Corresponding author email: vincent.maurice@chimie-paristech.fr

[§] Corresponding author email: philippe.marcus@chimie-paristech.fr

## 1. **Introduction**

The durable use of stainless steels (SS) in our daily lives and in many industries relies on self-protection against corrosion provided by a surface barrier oxide film, the passive film. Although a few nanometers in thickness, this film effectively isolates the metallic alloy substrate from the aggressive environment and thus provides long-term protection. In severe conditions, e.g. in chloride-containing environments, the stability of the passive film can fail locally and lead, in the absence of self-healing, to localized corrosion with costly consequences and environment- and even health-jeopardizing risks.

It is now well-established from surface analytical studies of passivity performed on ferritic and austenitic SS at macroscopic level that the key for efficient passivity is a marked Cr(III) enrichment of the passive film, because of the higher stability of Cr(III) compared to Fe(II,III) oxide/hydroxide species [1-8]. For austenitic SS, only very little Ni(II) is present when detected in the passive film and the metallic alloy region underneath the oxide is Ni(0)-enriched [9-15]. Most recently, it has been shown that the pre-passivation mechanism by which the surface oxide film initially grows can produce Cr(III)-depleted chemical/structural heterogeneities/defects at the nanometric scale [16-18]. Nanoscale compositional heterogeneities were also inferred from studies of mature passive films [15,19] and can be considered as weak points at the origin of the subsequent loss of stability [7,8]. Improving the stability of the corrosion protection requires thus to deeper understand the nanoscale pre-passivation mechanisms governing the chromium enrichment, like studied on single crystals starting from oxide-free surfaces [16-18]. This also requires to develop the knowledge of the alterations brought by exposure to aqueous environments and electrochemical passivation of oxide pre-covered SS surfaces, like addressed on polycrystalline samples starting from native oxide-covered surfaces [20-22].

In this work, we interrogated the compositional and nanoscale morphological modifications brought by exposing metallic and oxide pre-covered austenitic SS surfaces to water-containing gaseous and liquid environments and by passivation in aqueous acid solution. Complementary X-ray

Photolectron Spectroscopy (XPS) and Scanning Tunneling Microscopy (STM) surface analysis was applied. We used a closed system with direct connection between an ultra-high vacuum (UHV) platform for surface preparation and analysis and an Ar-filled glove box for electrochemical treatment in order to avoid the modifications brought by transfer in ambient air. Such a closed system also allows to test different well-controlled pre-oxidation scenarios starting from an oxide-free surface state. A (100)-oriented Fe-18Cr-13Ni(at%) single crystal was used as a model of the most common AISI 304 SS grade. Single crystals have seldom been used compared to polycrystals to study surface reactivity and passivation of stainless steel [14-18,23-30] despite the fact that they allow deeper mechanistic insight by avoiding the effects of multiple crystalline orientations and microstructural defects. Moreover, an adequate preparation under UHV yields atomically flat metallic surfaces which is a prerequisite to interrogate the topographic and structural surface alterations generated at the nanoscale by the growth of ultra-thin surface oxide films and their subsequent modifications. Here oxide-free and well-ordered atomically flat surfaces were prepared in the UHV platform and directly transferred without exposure to air to the glove box under argon atmosphere and equipped for electrochemistry. The results obtained using such a direct transfer line bring new insight on how the pre-oxidized surface is modified upon electrochemical passivation and on the effects of different pre-oxidation scenarios starting from a well-defined initial oxide-free surface state. They show that controlling the pre-oxidation conditions can promote the protectiveness and stability of the passive film on stainless steel.

## 2. Experimental

Experiments were performed in a closed system combining a UHV platform for surface preparation and analysis and a glove box equipped for electrochemistry with direct transfer of samples from UHV gaseous environment to liquid environment without exposure to ambient air (Figure 1). The UHV platform consists of a preparation chamber, an XPS chamber and an STM chamber. The base pressure is lower than $10^{-10}$ mbar. The glove box is under Ar atmosphere (purity 99.998%, overpressure with respect to atmospheric pressure $\Delta P = 215$ Pa). The typical residual concentrations of $O_2$ and $H_2O$ are about 350 and 150 ppm, respectively, in the presence of aqueous solutions. The corresponding relative humidity is 0.7-0.5% at 20-25°C. Surface cleaning, pre-oxidation and characterization by XPS and STM were performed in the UHV platform, while the exposure to liquid environments and electrochemical treatments were performed in the directly connected glove box.

*2.1. Sample preparation*

A (100)-oriented single crystal of Fe-18Cr-13Ni (at.%) composition (purity 99.999%) purchased from MaTeck was used. The surface was first prepared by mechanical and electrochemical polishing as detailed elsewhere [14]. After introduction of the sample in the UHV platform, the surface was further treated by cycles of $Ar^+$ ion sputtering (1 kV, 10 µA, 10 minutes) and annealing (800°C, 10 minutes) in the preparation chamber, in order to obtain a clean and well-ordered surface at atomic level as routinely verified by XPS and Low Energy Electron Diffraction (LEED), respectively. Hereafter we refer to this sputter-cleaned and annealed surface as to initial oxide-free surface. All oxidation- and passivation-induced changes were studied starting from this initial state.

The growth of the surface oxide film can be controlled in the UHV system by well-defined exposure to gaseous oxygen at ultra-low pressure (ULP). The exposure is directly obtained by integration of recorded oxygen pressure as a function of exposure time. Our previous study of initial oxidation in these conditions showed that the surface is already saturated at room temperature (RT) after an oxygen exposure of 10 L ($1 \text{ L} = 1.33\times10^{-6}$ mbar × 1 s) [16,17]. Here it was chosen to expose the initial oxide-free surface to gaseous oxygen up to 100 L in the preparation chamber to ensure saturation of the oxide growth at RT. This as-prepared surface is referred to as ULP pre-oxidized surface.

For exposure to liquid aqueous environments and electrochemical treatments, the surfaces prepared under UHV were transferred to electrochemical cells under the Ar environment of the glove box at RT. This environment contains a residual of gaseous oxygen and water vapor that causes oxidation. The initial oxide-free surface directly exposed to the glove box environment is referred to as GB pre-oxidized surface, a designation only used in Tables and Figures. Two durations of pre-oxidation in the glove box were tested on the oxide-free surface in order to determine the condition needed to reach saturation of the oxide growth. No oxidation kinetics was recorded. This surface was further exposed to aqueous sulfuric acid for passivation. The ULP pre-oxidized surface being already saturated by a surface oxide film, its alterations were studied for exposure to ambient air and to liquid ultra-pure water before passivation in the aqueous sulfuric acid environment.

*2.2. Electrochemical measurements*

For exposure to liquid aqueous environments and electrochemical treatments, we used a micro electrochemical cell controlled by a PicoStat potentiostat and Picoscan software (Agilent Technologies). This cell, also used for STM measurements under electrochemical control, and its cleaning process have been detailed elsewhere [30]. The electrochemical cell is made of Kel-F and contains ~350 µl of electrolyte. A Viton O-ring delimits a working electrode area of 0.16 $cm^2$. This working area is well adapted for centering the cell on the surface of the SS sample. During the electrochemical measurements, the sample can thus remain attached to the sample holder needed for transfer from/to UHV. Two Pt wires served as pseudo reference electrode (calibrated as E/SHE = E/Pt + 0.75 V) and counter electrode.

The electrolyte used for passivation was a 0.05 M $H_2SO_4$ aqueous solution prepared from ultrapure chemicals (VWR®) and Millipore® water (resistivity > 18 MΩ cm) and bubbled with argon for 30 minutes prior to introduction in the glove box. Polarization curves were obtained by linear scan voltammetry performed in the range of (-1.2 V, +1.2 V) at a scan rate of 5 mV/s after resting at open circuit potential (OCP) for 30 minutes. Passivation treatments were conducted at free potential (i.e. at OCP) for 30 minutes and with anodic polarization applied by stepping the potential to -0.1 V/Pt for 30 minutes after resting at OCP for 30 minutes. The effects of treatment time at OCP and under anodic polarization are beyond the scope of the present study. No cathodic pre-treatment was performed in order to avoid any reduction-induced alteration of the surface oxide films prior to passivation. Exposures to liquid water were performed using the same micro cell and Millipore® water at free potential for 30 minutes. All experiments were performed at room temperature.

### *2.3. Surface characterization*

Chemical composition and nanoscale morphology of the oxide-free, ULP and glove box pre-oxidized surfaces, and of the alterations produced by exposure to air, liquid water and sulfuric acid environments were examined under UHV by XPS and STM, respectively. XPS analysis was performed with an Argus spectrometer using the monochromatic XM1000 MkII Al $K_{\alpha}$ radiation (1486.6 eV), both from Scienta Omicron. Pass energy was 20 eV for high resolution spectra. Take-off angle of the photoelectrons was 45°. The spectra were decomposed with CasaXPS [31] using non-linear background and asymmetric line shape for the metallic elements. The spectral decomposition method adopted in our previous study was applied [16]. STM experiments were conducted with a VT STM XA from Scienta Omicron. Electrochemically etched tungsten tips were used and prepared under UHV by heating, voltage pulses and high voltage scanning. Images were corrected from the slope of the terraces. Image parameters, such as brightness, contrast and saturation, were adjusted to present more details. No image filtration was applied.

## 3. Results and discussion

### *3.1. Electrochemical passivation in sulfuric acid electrolyte*

Figure 2 compares the potentiodynamic polarization curves obtained in 0.05 M $H_2SO_4$ for the GB (glove box) pre-oxidized and ULP pre-oxidized samples. The two pre-oxidized surfaces have the same corrosion potential of -0.92 V/Pt but both the cathodic and anodic branches show lower currents and thus reactivity for the ULP pre-oxidized surface. In the anodic branch, the current density of the ULP pre-oxidized sample is lower prior to passivation and at the passivation potential, indicating that the oxide film formed under ULP conditions and exposed to the acid solution at open circuit potential is more protective against active dissolution than the oxide film directly formed in the glove box environment and exposed to the acid solution in the same conditions. In the passive domain, a clear difference subsists with a lower current density still observed for the ULP pre-oxidized sample and showing the formation of the better protective passive film after anodic passivation. The entry in the transpassive domain is also retarded for the ULP pre-oxidized sample showing a beneficial effect of ULP pre-oxidation on the stability of the passive film formed in the acid electrolyte. The passive film remains also more protective in the transpassive domain for the ULP pre-oxidized sample.

### *3.2. Composition of pre-oxidized and passivated surfaces*

Figure 3 shows the XPS Fe 2p, Cr 2p and Ni $2p_{3/2}$ core level spectra and their reconstruction for the surfaces pre-oxidized and passivated in the different conditions studied in this work. Figure 3 (I) presents the spectra for the initial oxide-free surface. Their envelopes were taken as references for the metallic components of the pre-oxidized and passivated surfaces. Starting from this initial oxide-free surface, the spectra of the glove box pre-oxidized surfaces obtained after 5 and 30 minutes are shown in Figure 3 (II) and (III), respectively. The glove box pre-oxidized surface was then passivated in 0.05 M $H_2SO_4$ at OCP (Figure 3 (IV)). Figure 3 (V) shows the spectra of the ULP pre-oxidized surface prepared under UHV from the initial oxide-free surface. This ULP pre-oxidized surface was exposed to air for 10 minutes (Figure 3 (VI)), in order to be compared with the glove box pre-oxidized surfaces. Then, the ULP pre-oxidized surface was exposed to ultra-pure liquid water (Figure 3 (VII)) and passivated in 0.05 M $H_2SO_4$ at OCP (Figure 3 (VIII)) and at −0.1 V/Pt (Figure 3 (IX)).

Table 1 compiles the curve fitting parameters used to reconstruct the Fe 2p, Cr 2p and Ni $2p_{3/2}$ core level spectra. They differ slightly from those used to fit the core level spectra obtained in the same analytical conditions in our previous study of the initial oxidation of the Fe-18Cr-13Ni(100) single crystal surface [16]. This is because the changes of environment for pre-oxidation and passivation induce different chemical shifts and full widths at half maximum. These parameters were fixed for all the spectra analyzed in this work. Only the intensities of the peaks were adjusted for optimizing the reconstruction. Table 2 compiles the resulting intensity values converted in relative atomic concentrations values for all components by taking into account the transmission factors of the analyzer and the photoionization cross sections of the elements. Such a calculation inherently assumes a homogenous depth distribution of the identified surface species. This first estimate of the surface composition is further on completed with values obtained using layered models.

The Cr 2p spectra show the presence of $Cr^0$ metal, Cr-N, $Cr^{3+}$ oxide and $Cr^{3+}$ hydroxide surface species. The co-segregation of Cr and N induced by annealing under UHV causes the formation of Cr-N on the initial oxide-free surface (Figure 3, Table 2 (I)), confirming our previous study [16]. Cr-N remains observed after ULP pre-oxidation (Figure 3, Table 2 (V)) and subsequent treatments in air and liquid water (Figure 3, Table 2 (VI) and (VII)). The $Cr^{3+}$ oxide component is identical to that measured in our previous work [16]. The additional component assigned to $Cr^{3+}$ hydroxide is pronounced after passivation in $H_2SO_4$ (Figure 3, Table 2 (IV), (VIII) and (IX)). This additional component is shifted by +0.7 eV compared to the $Cr^{3+}$ oxide component. This chemical shift is consistent with the binding energy reported in the literature for $Cr^{3+}$ hydroxides [13,14,20,21,22,24,32,33].

The Fe 2p spectra show the presence of the $Fe^0$, $Fe^{2+}$ and $Fe^{3+}$ chemical states. No obvious chemical shift indicative of the variation of the chemical environment of the oxidized states, such as oxide, spinel or hydroxide matrices, could be identified. The main iron oxidized state is $Fe^{3+}$. $Fe^{2+}$ is only observed at few atomic percent in most cases (Table 2). No $Ni^{2+}$ oxidized state was observed after ULP pre-oxidation under UHV. It only appears in minute amounts in the form of oxide and hydroxide species after exposure at atmospheric pressure in the glove box environment and in ambient air and after anodic passivation (Table 2).

The O 1s core level spectra for pre-oxidation and passivation are presented in the Figure 4 and Figure 5, respectively. They show the presence of oxide ($O^{2-}$), hydroxide ($OH^-$) and adsorbed water/oxygen ($H_2O$) components and their variations for different surface treatments. For the pre-oxidized surfaces prepared without contact with aqueous electrolyte (Figure 4), the dominant species are the oxide ligands ($O^{2-}$), while for the passivated surfaces (Figure 5) the dominant species are the hydroxide ligands ($OH^-$). Water ligands are only observed for the passivated surfaces and more water is bound after OCP treatment of the ULP pre-oxidized surface.

### 3.2.1. Glove box pre-oxidation and passivation-induced alterations

Table 2 shows that the two surfaces pre-oxidized in the glove box environment for 5 and 30 minutes have very similar compositions, as expected from the very similar shape of the core level spectra (Figure 3 (II) and (III)). The proportions of each element in the metallic and oxidized states as well as those of the oxide components are identical considering the uncertainty of the measurement, showing that most surface alterations associated with the formation of the native oxide film in the glove box environment occurred already after an exposure of 5 minutes. As expected, the oxide films are enriched in chromium and the metallic phases appear depleted in metallic chromium compared to the initial oxide-free surface. $Cr^{3+}$ hydroxide is detected in very low amount only after the 30-minute exposure. About 90% of oxidized iron are $Fe^{3+}$ species. As indicated above, no chemical shift of the iron oxide components was observed because the binding energies of iron oxides are too close to those of iron hydroxides or oxyhydroxides [14,32-35]. However, the O 1s core level spectrum shows the presence of $OH^-$ ligands (Figure 4), thus indicating that a significant fraction of $Fe^{2+,3+}$ species are in the form of hydroxides or oxyhydroxides. $Ni^{2+}$ oxide and hydroxide species are observed in very low amounts and the metallic phase appears enriched in nickel compared to the initial oxide-free surface.

After exposure to sulfuric acid at OCP (Figure 3 (IV)), the oxide becomes composed of 73% $Cr^{3+}$ species, of which 73% are hydroxides (Table 2). This markedly increased enrichment in $Cr^{3+}$ compared to the glove box pre-oxidized state is attributed to the preferential dissolution of the $Fe^{2+,3+}$ species in acid solution [36,37], and confirms previous observations with a direct transfer line avoiding ambient air re-oxidation [21,38]. The O 1s core level spectrum shows that $OH^-$ groups become the dominating ligands (Figure 5), in agreement with the marked enrichment in $Cr^{3+}$ hydroxides. The remaining $Fe^{2+,3+}$ species are in the form of oxides or hydroxides (or oxyhydroxides). Thus, while iron (hydr)oxides dissolve in contact with the acid electrolyte, chromium hydroxides preferentially form in the oxide film. Their formation probably includes the hydroxylation of the chromium oxide species already present after pre-oxidation since the composition of the metallic phase does not indicate further preferential consumption of metallic chromium. Due to the preferential consumption and dissolution of iron, the proportion of metallic nickel increases. Since partially dissolving at OCP, a decrease of the thickness of the oxide film can be expected [21,38]. This is supported by the decrease of the total concentration of oxidized species (40% after passivation vs. ~55% prior to passivation at OCP) and confirmed by the layered models discussed further on.

#### 3.2.2. ULP pre-oxidation and water-induced alterations

The ULP pre-oxidized surface (Figure 3 (V)) contains less oxidized (34%) and more metallic (66%) species compared to the glove box pre-oxidized surface (~55% and ~45%, respectively) (Table 2), suggesting the formation of a thinner oxide film also confirmed further on. Regarding the relative concentration in oxidized chromium and iron, the ULP pre-oxidized surface contains more chromium oxide (56%) and less iron (hydr)oxide (38%) species compared to the glove box pre-oxidized surface (33-35% and 61-64%, respectively). The formation of chromium oxide is favored under ULP oxygen gas and leads to a markedly more Cr-enriched oxide film, owing to the thermodynamically favored oxidation of Cr as previously confirmed [16,17]. This difference can be explained by the much larger residual of water vapor in the glove box environment than in the ULP environment. The water vapor promotes the formation of iron (hydr)oxide species in the glove box environment, in agreement with the acceleration effect observed on the oxidation behavior of SS at high temperature but also at RT [39]. The formation of more hydroxylated oxide films in the glove box is supported by the line shape of the O 1s core level spectrum (Figure 4). The metallic phase of the ULP pre-oxidized surface appears also depleted in chromium compared to the initial oxide-free surface owing to the preferential consumption of this element by the pre-oxidation treatment. However, it does not appear enriched in nickel despite the absence of oxidation of this element. The presence of chromium nitride resulting from co-segregation of Cr and N during surface preparation of the initial oxide-free surface subsists after ULP pre-oxidation, like previously observed [16].

After exposure to ambient air (Figure 3 (VI)), the ULP pre-oxidized surface is further oxidized as shown by the increase and decrease of the concentrations in oxidized (54% vs 34%) and metallic (46% vs 66%) species, respectively (Table 2). Iron oxidizes more rapidly than chromium, markedly decreasing the enrichment in $Cr^{3+}$ oxide. Nickel also oxidizes very weakly, unlike under ULP environment but similar to under glove box environment. The water vapor in the air favors mainly the oxidation of iron [39] and triggers that of nickel, like in the glove box environment also containing a water vapor residual. Compared to the glove box pre-oxidized surface, the ULP pre-oxidized surface subsequently exposed to air has a similar composition but is slightly more enriched in $Cr^{3+}$ oxide (39%) than the glove box pre-oxidized surface (33-35%), due to the high initial $Cr^{3+}$ enrichment (56%) provided by the ULP pre-oxidation treatment and that partially persists. In addition, the ULP pre-oxidized surface subsequently exposed to air appears to contain more metallic iron than the glove box pre-oxidized surface, suggesting that the ULP pre-formed oxide film better inhibits iron oxidation in the tested conditions of exposure to air.

After exposure in liquid water (Figure 3 (VII)), the ULP pre-oxidized surface shows a similar but more marked evolution of the composition than after exposure to ambient air. It is further oxidized as indicated by the further increase and decrease of the concentrations in oxidized (58% vs 54%) and metallic (42% vs 46%) species, respectively (Table 2). The $Cr^{3+}$ oxide enrichment is further decreased (29% vs 39%) due to the preferential oxidation of iron corroborated by the diminution of metallic iron (62% vs 70%). Chromium nitride is still observed and its proportion decreases. No chromium hydroxide was observed despite exposure in liquid water. These observations suggest that the dominant reaction remains iron oxidation and that iron (hyd)oxide species form at the surface of the pre-formed Cr-enriched oxide layer.

This comparative analysis shows a more pronounced enrichment in $Cr^{3+}$ of the oxide film formed by pre-oxidation under ULP gaseous oxygen. Upon contact with water vapor, the $Cr^{3+}$ enrichment is attenuated by competitive iron oxidation but partially preserved. In liquid water, the initial further $Cr^{3+}$ enrichment provided by ULP pre-oxidation becomes fully compensated by preferential iron oxidation. The persistence of chromium nitride pre-formed on the initial oxide-free surface after ULP pre-oxidation and subsequent exposure to water vapor is consistent with the protective nature provided by the ULP pre-oxidation treatment.

3.2.3. Effect of ULP pre-oxidation on passivation-induced alterations

After exposure to sulfuric acid at OCP of the ULP pre-oxidized surface (Figure 3 (VIII)), the oxide is composed of 71% $Cr^{3+}$ species, of which 93% are hydroxides (Table 2). The composition is very close to that of the glove box pre-oxidized surface treated in the same conditions, except for a higher ratio in $Cr^{3+}$ hydroxides species. Note that these are the two surfaces compared in the electrochemical measurements discussed above (Figure 2), showing a beneficial effect of the ULP pre-oxidation treatment on the corrosion resistance. Compared to the ULP pre-oxidized surface subsequently exposed to air and to liquid water, the surface film formed at OCP contains markedly less iron (hydr)oxides which are dissolved like observed for the glove box pre-oxidzed surface. This observation is made possible by the direct transfer line that avoids exposure to ambient air. Oxide dissolution is supported by the decrease and increase of the concentrations in oxidized (38% vs 58%) and metallic (62% vs 42%) species, respectively. It is also shown by the layered models discussed thereafter. After exposure to sulfuric acid at OCP, the predominant $Cr^{3+}$ species become hydroxides also like observed for the glove box pre-oxidized surface and confirming results obtained on polycrystalline 316L SS samples also with a direct transfer line [21]. The $Cr^{3+}$ enriched oxide layer formed by ULP pre-oxidation is likely involved in the conversion of chromium oxides to chromium hydroxides due to the preferential dissolution of oxidized iron. This is also supported by the layered models discussed below. In the metallic phase, the proportions of chromium and nickel increase after treatment at OCP, due to the preferential consumption of metallic iron, like observed on the glove box pre-oxidized surface.

After anodic passivation in sulfuric acid (Figure 3 (IX)), one observes the increase and decrease of the concentrations in oxidized (59% vs 38%) and metallic (41% vs 62%) species, respectively (Table 2). This is consistent with the re-growth of the oxide film partially dissolved at OCP [21,38]. The oxide is composed of 74% $Cr^{3+}$ species, of which 39% are hydroxides. Oxide becomes the predominant $Cr^{3+}$ species as a result of the reaction of dehydroxylation promoted by anodic polarization in the passive domain, like previously reported [14,15,20,21,24,38,40]. A very small amount of oxidized nickel ($Ni^{2+}$) is detected in the oxide after anodic polarization. In the metallic phase, iron is preferentially consumed owing to dissolution of its (hydr)oxides and nickel becomes markedly enriched due to preferential consumption of iron and chromium by the re-growth of the passive film.

### 3.2.4. Layered oxide film models

Layered models are often used in the literature to calculate oxide film thickness and composition from XPS data based on the exponential decrease of the photoelectron intensity with increasing depth of emission from the topmost surface plane [13,14,20-24,40,41-45]. Here we have applied two types of models in order to discuss the passive layer chemical structure: a single layer model (Figure 6(a)) and a bilayer model (Figure 6(b)).

The single layer model assumes that all oxide and hydroxide components originate from a single surface oxide layer. For the calculation, we considered that $Fe_2O_3$, $Cr_2O_3$ and $Cr(OH)_3$ are the constitutive phases based on our measurements of $Fe^{3+}$ as largely dominant iron oxide species and $Cr^{3+}$ as chromium oxide and hydroxide species. The bilayer model assumes an outer layer of exchange with the electrolyte and an inner layer being the barrier slowing down atomic transport. We considered an outer layer constituted of $Fe_2O_3$ and $Cr(OH)_3$ and an inner layer of $Cr_2O_3$ based on bilayer structures observed previously [4,5,6,13-15,20-22,24,38,40]. The intensities of the $Fe^{2+,3+}$ oxide, $Cr^{3+}$ oxide and $Cr^{3+}$ hydroxide components were assigned consistently. The low intensity $Ni^{2+}$ components were neglected. For both models, the intensities of the $Fe^0$, $Cr^0$ and $Ni^0$ components were assigned to a metallic phase underneath the oxide film of composition modified with respect to bulk composition. In addition, such multilayer models assume that the different layers are homogeneous in thickness and separated by sharp interfaces, that the oxide or metal components are homogeneously distributed in each layer, and that the metal signal is totally derived from the region underlying the oxide. It is therefore the equivalent thickness of the oxide layers that is calculated. Calculations were carried out using intensity ratios of the (hydr)oxide to metal components.

The results obtained with the single layer model are compiled in Table 3. The thickness values of the oxide layer for the glove box- and air-formed films and for the passive films formed by anodic polarization ranges from 1.2 to 1.6 nm, which is in the lower range of the typical thickness of 1−3 nm for room temperature native oxide films and passive films on stainless steels [11-15,20,21,22,24,38,40,44-47]. The oxide film formed in the glove box environment has the same thickness (1.3-1.4 nm) after 5 or 30 minutes but becomes slightly further enriched in chromium as a result of chromium hydroxide formation after prolonged exposure. This shows that an exposure beyond 5 minutes has no effect on the thickness and only a slight effect on the composition of the native oxide film formed in glove box environment. After exposure to the acid solution at OCP, the oxide film markedly decreases in thickness to 0.8 nm and becomes markedly further enriched in chromium as a result of the preferential dissolution of the iron (hydr)oxide species, thus confirming our analysis above and the result obtained with the direct transfer line.

The oxide film formed by ULP pre-oxidation is found thinner and more enriched in chromium that those formed by glove box pre-oxidation. This is consistent with the presence of a residual of water vapor in the glove box environment promoting the competitive oxidation of iron and thus the formation of a thicker film [39]. After exposure in ambient air and liquid water, the oxide film becomes thicker and less enriched in chromium. This confirms that the interaction with water, in vapor or liquid form, promotes the competitive oxidation of iron and the resulting film thickening. The oxide film formed after exposure to liquid water has a similar thickness and composition as that formed by glove box pre-oxidation despite the initial much higher Cr enrichment brought by the ULP pre-oxidation treatment. After treatment in the acid solution at OCP, the same trends, i.e.

thickness decrease and marked increase of the Cr enrichment, both due to preferential iron (hydr)oxide dissolution, are observed as for the glove box pre-oxidized surface. Thickness and composition of the oxide film are then identical with no apparent remaining effect of the initial much higher Cr enrichment. The higher relative concentration in chromium hydroxide brought by ULP pre-oxidation is not accounted for in this single layer model analysis. Passivation in the acid solution with applied anodic polarization causes the film to re-grow and become further enriched in chromium, confirming our analysis above. This is assigned to competitive iron and chromium oxidation combined with preferential iron (hydr)oxide dissolution, like previously observed on 316L SS also thanks to a direct transfer procedure [21].

For both the glove box and ULP pre-oxidized surfaces, the variations of composition in the modified metallic alloy underneath the oxide film after further treatment are not strong but are consistent with the variations of the Cr enrichment in the oxide film.

The results obtained with the bilayer model, compiled in Table 4, confirm the above trends and bring further insight. The total thickness of the glove box- and air-formed films and of the passive films formed by anodic polarization ranges slightly higher, from 1.7 to 2.2 nm, in better agreement with literature values also calculated with bilayer models [11-15,20-22,24,38,40,44]. The variations induced by the surface treatment confirm the trends obtained with the single layer model. The same thickness of the 5- and 30-minute glove box pre-oxidized surfaces is confirmed but the slight increase of the Cr enrichment due to chromium hydroxide formation after prolonged exposure is not significant with this model. After treatment in the acid solution at OCP, the thickness decrease is confirmed but it is calculated to mostly occur in the inner layer whereas dissolution of iron (hydr)oxide species is expected to mostly affect the outer layer. This inconsistency is only apparent. It arises from the formation of chromium hydroxide, assigned in the model to the outer layer and thus compensating the decrease of the outer layer thickness. This result of the bilayer model is consistent with chromium oxide initially in the inner layer being transformed into chromium hydroxide, thus decreasing the inner layer thickness and resulting in an outer layer markedly enriched in chromium.

The oxide film formed by ULP pre-oxidation is also found thinner and more enriched in chromium that those formed by glove box pre-oxidation. After exposure in ambient air and liquid water, the oxide film increasingly thickens and the chromium content increasingly decreases due to preferential iron oxidation, as reflected by the increase of the outer layer thickness. The oxide film formed after exposure in liquid water is slightly thicker and less Cr-enriched than that formed by glove box pre-oxidation despite the initial higher Cr enrichment. After treatment in the acid solution at OCP, the decrease of the inner layer thickness observed for the glove box pre-oxidized surfaces is reproduced, which can also be explained by the transformation of chromium oxide into chromium hydroxide concomitantly to the dissolution of the iron (hydr)oxide species. The higher content in chromium hydroxide compared to the glove box pre-oxidized surface treated at OCP is reflected by the slight increase in thickness and slightly higher chromium concentration of the outer layer. After passivation in the acid solution with applied anodic polarization, the film re-growth is observed in the inner layer as result of dehydroxylation of chromium hydroxide promoted by anodic polarization. The competitive iron oxidation mitigates the chromium enrichment in the outer layer.

The calculated values of the composition in the modified metallic alloy underneath the oxide film are identical to those obtained with the single layer model taking into account the uncertainty.

### 3.2.5. Summary of the XPS analysis

The XPS analysis presented above shows that the oxide films formed by exposure to the environment of the glove box, to ambient air and to liquid water are less enriched in chromium because of the enhanced oxidation of iron in the presence of water in vapor or liquid form [39]. Oxidized chromium is mostly present as $Cr^{3+}$ oxide whereas oxidized iron (essentially $Fe^{3+}$) is present as oxide, hydroxide or oxyhydroxide species that could not be differentiated. Quantification of the data shows only slight differences of composition and thickness despite the different preparation conditions. The pronounced effect of the ULP pre-oxidation on the $Cr^{3+}$ enrichment does not persist in contact with pure liquid water. The conditions of exposure to $H_2O$ are too far beyond the initial stages of interaction and the surface is too quickly saturated to capture accurately the effect of ULP pre-oxidation on the reactivity towards $H_2O$. Exposing oxide-free and oxide pre-covered surfaces to water vapor in ULP conditions should bring further insight into the detailed mechanisms of initial oxidation of iron and chromium on stainless steel. The chromium nitride formed by co-segregation on the initial oxide-free surface remains after exposure to air and liquid water of the ULP pre-oxidized surface whereas it is decomposed on the glove box pre-oxidized surface, thus showing that the oxide film formed by ULP pre-oxidation would protect this species from the decomposition induced by the direct contact with water vapor and dioxygen in the environment of the glove box. This is consistent with this Cr-enriched surface oxide film forming a barrier preventing/retarding subsequent oxidation (mostly of iron) as concluded from the study of the initial oxidation stages [16,17] and confirmed here after exposure to ambient air of the ULP pre-oxidized surface.

After immersion in acid solution, the XPS analysis also shows further $Cr^{3+}$ enrichment due to preferential dissolution of the $Fe^{3+}$ (hydr)oxide species with transformation of the exposed $Cr^{3+}$ oxide species into $Cr^{3+}$ hydroxide species. Anodic polarization forces the re-growth of the oxide film with $Cr^{3+}$ re-enrichment due to preferential iron (hydr)oxide dissolution and dehydroxylation of the chromium hydroxide species. The use of the direct line transfer between electrochemical treatment and surface analysis was decisive in revealing these modifications of the surface oxide film as re-oxidation in ambient air can be expected to modify the films, especially after OCP treatment where the oxide is partially dissolved and thinner.

The significant effect of ULP pre-oxidation on the composition and thickness of the surface oxide films after exposure in the acid solution at open circuit potential is the increase of the relative concentration in chromium hydroxide, reflected in the bilayer model analysis by the slight increases in thickness and Cr concentration of the outer layer of the oxide film. More bound water is also observed after ULP pre-oxidation. Global oxide film thickness is not changed significantly after

OCP treatment and cannot be advanced to explain the improved corrosion resistance provided by ULP pre-oxidation. Coupled with alterations of the morphology and/or nanostructure of the surface oxide film, the increased relative concentration in chromium hydroxide may explain the improved protectiveness and stability of the passive film provided by the ULP pre-oxidation treatment. An effect of increased bound water cannot be excluded.

*3.3. Nanoscale morphology of pre-oxidized and passivated surfaces*

Figure 7 shows the nanoscale morphology of the initial oxide-free surface as measured under UHV by STM. Two types of regions were observed, exposing (i) narrower terraces separated by atomic multi steps or by the accumulation of emerging screw dislocations (Figure 7 (a)) or (ii) extended terraces with a lower density of emerging screw dislocations (Figure 7 (b)). The annealing process causes the formation of an ordered atomic superstructure in the topmost atomic plane terminating the terraces. This superstructure, denoted $(\sqrt{2}\times\sqrt{2})R45°$ and described in details elsewhere [18], was confirmed but is not presented here. At the nanometric scale, one also notices the formation of islands of orthogonal shape formed on the narrow and on the extended terraces. Their size ranges between ∼10 nm and ∼200 nm and their apparent height is from 30 to 100 pm above the surrounding terrace. These islands are ordered with the formation of a self-organized structure that will be discussed in a separate article. As discussed previously [16] and confirmed in the present study by XPS, chromium nitride surface species were formed and observed to be accumulated in the terraces borders adjacent to the multi steps [18]. In the present work, they were not found to have any significant impact on the morphology of the surfaces after saturation by the 3D oxide films formed by oxidation and modified by passivation.

Figure 8 (a) shows that the pre-oxidized surface obtained by direct exposure to the glove box environment still displays regions with narrower or extended terraces and emerging screw dislocations. Darker appearing islands, with a difference in apparent height of ~60 pm, are observed on the terraces (Figure 8 (a,b)). They are preferentially located at the termination of the emerging screw dislocations. Their presence could be associated with the self-organized islands also observed, for some of them, at the termination of screw dislocations on the initial oxide-free surface and thus leaving imprints after pre-oxidation. The 3D oxide film completely covers the surface with a nanogranular morphology, homogeneous inside and outside the darker appearing islands (Figure 8 (b)). The size of the oxide grains measured at the topmost surface of the oxide film is estimated to ~3 nm.

After exposure to liquid water, that causes preferential further growth of iron (hydr)oxide species as shown by XPS, the terraces and the multi steps remain observable but are less well-defined (Figure 8 (c)). The surface is covered by a nanogranular film, but only partially in its outer part (Figure 8 (d)). The resulting morphology appears rough and nanoporous, which is assigned to the non-homogeneous formation of iron (hydr)oxides. The size of the oxide grains appear larger, about 10 nm, than prior to exposure to liquid water. Unlike after pre-oxidation, there are no remaining traces of the self-organized islands present on the oxide-free surface, which is assigned to further growth of the oxide film. After immersion in 0.05 M $H_2SO_4$ at OCP, that causes preferential dissolution of the iron (hydr)oxide species as shown by XPS, the terraces and multi steps are better defined again (Figure 8 (e)). The nanogranular morphology of the 3D oxide layer is more homogeneous and less porous (Figure 8 (f)). The size of the oxide grains is estimated to ~5 nm. This more compact nanogranular morphology is assigned to the markedly more Cr-enriched and thinner oxide film formed by passivation at the open circuit potential.

Figure 9 (a,b) shows the surface morphology obtained after ULP pre-oxidation and immersion in ultra-pure water. Like for the glove box pre-oxidized surface also immersed in liquid water, the terraces and multi steps are less well-defined (Figure 9 (a)), which can be assigned to the thickening of the oxide film resulting from the preferential further growth of iron (hydr)oxides. On the terraces (Figure 9 (b)), protruding islands are also observed, but in smaller number than on the glove box pre-oxidized surface exposed to water (Figure 8 (d)). After immersion in 0.05 M $H_2SO_4$ at OCP, causing preferential dissolution of the iron (hydr)oxides species, this ULP pre-oxidized surface also regains a more homogeneous morphology with better defined terraces and multi steps (Figure 9 (c)) and a more compact nanogranular morphology of the covering 3-dimensional (3D) oxide film (Figure 9 (d)). The oxide grains have a size estimated to ~2 nm, smaller than those formed without the ULP pre-oxidation treatment (~5 nm). After anodic passivation at −0.1 V, that causes re-growth of the oxide film and promotes its Cr-enrichment as measured by XPS, the ULP pre-oxidized surface displays less well-defined terraces and multi steps which is assigned to the thickening of the oxide film (Figure 9 (e)). The oxide film has an equally homogeneous nanogranular morphology (Figure 9 (f)). The size of the oxide grains is estimated to ~3 nm, slightly larger than that before polarization. This variation could result from the dehydroxylation and growth of $Cr^{3+}$ species observed by XPS.

Combined with the XPS data, these STM results show that a nanoporous layer of iron (hydr)oxides species is developed in the outer part of the surface oxide upon immersion in liquid water and that this layer is dissolved in the acid solution at open circuit potential. On the ULP pre-oxidized surface, protruding islands also grow in the outer part and but do not develop a nanoporous layer owing to hindered growth. They most probably consist of iron-rich (hydr)oxides particles since also dissolving in sulfuric acid. Their hindered growth is consistent with the markedly more Cr-enriched film formed by ULP pre-oxidation acting as a more effective barrier slowing down the formation of iron (hydr)oxides species in aqueous solution. After dissolution of the iron (hydr)oxides species and hydroxylation of the $Cr^{3+}$ species induced by immersion in acidic environment, the nanogranular oxide film is equally Cr-enriched but consists of smaller grains and is more hydroxylated than on the surface directly pre-oxidized in the glove box environment.

It can thus be concluded that pre-oxidation by exposure to gaseous oxygen at ultra-low pressure, by which a markedly more Cr-enriched barrier oxide layer is formed at the surface, prevents or delays the preferential reaction of iron in water-containing environment and promotes chromium hydroxide formation after dissolution of iron (hydr)oxides in acid solution. Increased chromium hydroxide formation and possibly bound water would be beneficial to the sealing of the weak sites of the barrier layer formed in acid solution at open circuit potential thus providing improved corrosion protection.

## 4. **Conclusion**

The passivation mechanisms and the effects of pre-oxidation were studied on (100)-oriented single crystalline model 304 austenitic stainless steel surfaces of Fe-18Cr-13Ni atomic composition by preventing contact to ambient air during transfer between ultra-low pressure environment for surface preparation (and analysis by XPS and STM) and aqueous environment for electrochemical treatment (and subsequent characterization). The use of a closed system with direct connection between UHV analytical platform and Ar-filled glove box equipped for electrochemistry was shown to be decisive in revealing the pre-oxidation effects and the electrochemically-induced alterations of the surface oxide films.

Starting from an oxide-free and atomically flat surface state prepared under UHV, pre-oxidation in dioxygen gas at ultra-low pressure and room temperature markedly promotes the Cr(III) enrichment of the oxide film grown at saturation. Without the ULP pre-oxidation treatment, the oxide film is only moderately enriched in Cr(III) and its barrier inner layer does not efficiently protect from the formation of iron (hydr)oxide species in $H_2O$-containing environments. An exchange outer layer of porous nanogranular morphology is formed at saturation of the reaction. With the ULP pre-oxidation treatment, the barrier inner layer of the oxide film prevents/delays subsequent preferential iron (hydr)oxides formation in the exchange outer layer.

Exposure to aqueous sulfuric acid solution at open circuit potential causes preferential dissolution of iron (hydr)oxide species exposing the Cr(III) oxide species of the barrier inner layer for hydroxide formation. With the ULP pre-oxidation treatment, Cr(III) hydroxide formation is promoted as well as compactness of the nanogranular morphology. Protection against active dissolution is improved. Anodic polarization in the passive domain forces the re-growth of the oxide film with Cr(III) further enrichment due to preferential iron (hydr)oxide dissolution. Cr(III) hydroxide species are dehydroxylated with the ULP pre-oxidation treatment promoting the protectiveness and stability of the passive film.

## Acknowledgments

This project has received funding from the European Research Council (ERC) under the European Union's Horizon 2020 research and innovation program (ERC Advanced Grant No. 741123, Corrosion Initiation Mechanisms at the Nanometric and Atomic Scales : CIMNAS). Région Île-de-France is acknowledged for partial funding of the XPS and STM equipment. China Scholarship Council (CSC) is acknowledged for the scholarship to the first author (No. 201606380129).

## Figure captions

*Figure 1 Closed system combining an ultra-low pressure (UHV) platform for surface preparation and analysis and a glove box equipped for electrochemistry with direct transfer of samples from UHV gaseous environment to liquid environment without exposure to ambient air. This apparatus is used to observe and characterize the modifications of metal and alloy surfaces caused by their interactions with the environment.*

*Figure 2 Potentiodynamic polarization curves for (100)-oriented Fe-18Cr-13Ni surfaces prepared oxide-free in UHV environment and subsequently pre-oxidized to saturation by exposure to dioxygen gas at ultra-low pressure (ULP) or directly in glove box (GB) environment. Recorded in 0.05 M* $H_2SO_4$ *at room temperature after 30 minutes rest at OCP with 5 mV/s scanning rate.*

*Figure 3 Fe 2p (a), Cr 2p (b) and Ni* $2p_{3/2}$ *(c) XPS spectra and peak fitting for (100)-oriented Fe-18Cr13Ni surfaces prepared oxide-free in UHV (I) and subsequently pre-oxidized directly in glove box (GB) environment for 5 minutes (II) and 30 minutes (III) before treatment in 0.05 M* $H_2SO_4$ *at open circuit potential (IV), and for surfaces pre-oxidized by exposure to ultra-low pressure (ULP) dioxygen gas (V) and then exposed to air (VI), to liquid water (VII) and treated in 0.05 M* $H_2SO_4$ *at open circuit potential (VIII) and passivated at − 0.1 V in the passive domain (IX). The Fe and Cr components are presented as sums of the main peaks and their satellite peaks. Each component is represented by a colored and shaded area; the envelope indicates the sum of all components; the dots represent the measured spectra. The emission angle is 45°.*

*Figure 4 Variation of O 1s XPS spectra for (100)-oriented Fe-18Cr13Ni surfaces in initial oxide-free state (I), in oxidized states obtained after 5-minute (II) and 30-minute (III) exposure in glove box (GB) environment, in oxidized states obtained by exposure to ultra-low pressure (ULP) dioxygen gas (V) followed by exposure to ambient air (VI).*

*Figure 5 Variation of O 1s XPS spectra for (100)-oriented Fe-18Cr13Ni surfaces in glove box (GB) pre-oxidized state passivated in 0.05 M* $H_2SO_4$ *at open circuit potential (IV) and in ultra-low pressure (ULP) pre-oxidized state passivated in 0.05 M* $H_2SO_4$ *at open circuit potential (VIII) and at − 0.1 V in passive domain (IX).*

*Figure 6 Single layer (a) and bilayer (b) models of surface oxide film formed on FeCrNi stainless steel. Alloy composition underneath the oxide film is modified compared to bulk composition.*

*Figure 7 Topographic STM images of oxide-free (100)-oriented Fe-18Cr-13Ni monocrystalline surface. (a) 400 nm×400 nm, I = 0.2 nA, V = 2.0 V. (b) 200 nm×200 nm, I = 0.5 nA, V = 2.0 V. Arrows point terminations of screw dislocations.*

*Figure 8 Topographic STM images of (100)-oriented Fe-18Cr-13Ni monocrystalline surfaces pre-oxidized in glove box environment (a,b) and exposed to liquid water (c,d) and subsequently to 0.05 M* $H_2SO_4$ *at OCP (e,f). (a,c,e) 1000 nm×1000 nm, (b,d,f) 200 nm×200 nm, (a) I = 0.2 nA, V = 2.0 V. (b) I = 0.5 nA, V = 2.0 V, (c-f) I = 0.5 nA, V = 0.5 V. Arrows point terminations of screw dislocations. Dashed circles mark nanopores in the outer layer of the oxide film.*

*Figure 9 Topographic STM images of (100)-oriented Fe-18Cr-13Ni monocrystalline surfaces pre-oxidized under UHV and exposed to liquid water (a,b) and subsequently to 0.05 M* $H_2SO_4$ *at OCP (c,d) and -0.1 V (e,f). (a,c,e) 1000 nm×1000 nm, (b,d,f) 200 nm×200 nm, (a-f) I = 0.5 nA, V = 0.5 V.*

# Tables

*Table 1 Curve fitting parameters used for reconstruction of XPS spectra*

| Core level | State | BE (±0.1 eV) | FWHM (±0.1 eV) |
|---|---|---|---|
| Fe $2p_{3/2}$ | $Fe^0$ | 706.8 | 0.8 |
| | $Fe^{2+}$ | 708.5 | 1.5 |
| | $Fe^{3+}$ | 710.5 | 3.1 |
| | $Fe^{2+}$ satellite | 713.0 | 4.0 |
| | $Fe^{3+}$ satellite | 716.2 | 4.3 |
| Fe $2p_{1/2}$ | $Fe^0$ | 719.9 | 1.6 |
| | $Fe^{2+}$ | 721.5 | 2.3 |
| | $Fe^{3+}$ | 723.8 | 3.6 |
| | $Fe^{2+}$ satellite | 726.5 | 4.0 |
| | $Fe^{3+}$ satellite | 729.9 | 4.3 |
| Cr $2p_{3/2}$ | $Cr^0$ | 573.8 | 1.1 |
| | Cr-N | 575.6 | 1.0 |
| | $Cr^{3+}$(oxide) | 576.4 | 2.0 |
| | $Cr^{3+}$(hydroxide) | 577.1 | 2.1 |
| | $Cr^{3+}$ satellite | 588.3 | 3.5 |
| Cr $2p_{1/2}$ | $Cr^0$ | 583.2 | 1.6 |
| | Cr-N | 585.1 | 1.8 |
| | $Cr^{3+}$(oxide) | 586.0 | 2.2 |
| | $Cr^{3+}$(hydroxide) | 586.7 | 2.2 |
| | $Cr^{3+}$ satellite | 597.3 | 4.8 |
| Ni $2p_{3/2}$ | $Ni^0$ | 852.8 | 0.9 |
| | $Ni^{2+}$(oxide) | 855.2 | 1.5 |
| | $Ni^{2+}$(hydroxide) | 856.2 | 1.7 |

*Table 2 Relative atomic concentrations (± 1 at%) of Fe, Cr and Ni in metallic and oxidized states for (100)-oriented Fe-18Cr-13Ni after different surface preparations.*

| No | Conditions | **Metal** | $Fe^0$ | $Cr^0$ | $Ni^0$ | **Oxide** | $Fe^{2+}$ | $Fe^{3+}$ | Cr-N | $Cr^{3+}$ (ox) | $Cr^{3+}$ (hyd) | $Ni^{2+}$ (ox) | $Ni^{2+}$ (hyd) |
|---|---|---|---|---|---|---|---|---|---|---|---|---|---|
| (I) | Oxide-free | **98** | 60 | 26 | 14 | **2** | 0 | 0 | 100 | 0 | 0 | 0 | 0 |
| (II) | GB pre-oxidized (5 min) | **45** | 62 | 20 | 18 | **55** | 5 | 58 | 0 | 33 | 0 | 2 | 2 |
| (III) | GB pre-oxidized (30 min) | **46** | 63 | 20 | 17 | **54** | 6 | 56 | 0 | 33 | 2 | 2 | 2 |
| (IV) | GB pre-oxidized + OCP exp. in acid | **60** | 54 | 23 | 23 | **40** | 5 | 23 | 0 | 20 | 53 | 0 | 0 |
| (V) | ULP pre-oxidized | **66** | 68 | 20 | 12 | **34** | 12 | 26 | 6 | 56 | 0 | 0 | 0 |
| (VI) | ULP pre-oxidized + exp. in air | **46** | 70 | 17 | 13 | **54** | 4 | 50 | 6 | 39 | 0 | 1 | 0 |
| (VII) | ULP pre-oxidized + exp. in water | **42** | 62 | 17 | 21 | **58** | 3 | 64 | 3 | 29 | 0 | 0 | 0 |
| (VIII) | ULP pre-oxidized + OCP exp. in acid | **62** | 55 | 21 | 24 | **38** | 3 | 26 | 0 | 5 | 66 | 0 | 0 |
| (IX) | ULP pre-oxidized + anodic pass. in acid | **41** | 51 | 20 | 29 | **59** | 3 | 20 | 0 | 46 | 29 | 1 | 1 |

*Table 3 Thickness and composition of oxide films and modified alloy regions as calculated using the single layer model for (100)-oriented Fe-18Cr-13Ni surfaces prepared as indicated.*

| No. | Conditions | Equivalent thickness *d* (± 0.1 nm) | Relative concentration (± 1 at%) | | | | |
|---|---|---|---|---|---|---|---|
| | | | Oxide layer | | Modified alloy | | |
| | | | Fe | Cr | Fe | Cr | Ni |
| (I) | Oxide-free | / | / | / | 76 | 17 | 7 |
| (II) | GB pre-oxidized (5 min) | 1.4 | 68 | 32 | 75 | 14 | 11 |
| (III) | GB pre-oxidized (30 min) | 1.3 | 63 | 37 | 75 | 15 | 10 |
| (IV) | GB pre-oxidized + OCP exp. in acid | 0.8 | 36 | 64 | 75 | 16 | 9 |
| (V) | ULP pre-oxidized | 0.7 | 47 | 53 | 81 | 13 | 6 |
| (VI) | ULP pre-oxidized + exp. in air | 1.2 | 57 | 43 | 79 | 14 | 7 |
| (VII) | ULP pre-oxidized + exp. in water | 1.5 | 69 | 31 | 75 | 16 | 9 |
| (VIII) | ULP pre-oxidized + OCP exp. in acid | 0.8 | 36 | 64 | 75 | 15 | 10 |
| (IX) | ULP pre-oxidized + anodic pass. in acid | 1.6 | 28 | 72 | 73 | 15 | 12 |

*Table 4 Thickness and composition of oxide films and modified alloy regions as calculated using the bilayer model for (100)-oriented Fe-18Cr-13Ni surfaces prepared as indicated.*

| No. | Conditions | Equivalent thickness (± 0.1 nm) | | | Relative concentration (± 1 at%) | | | | |
|---|---|---|---|---|---|---|---|---|---|
| | | | | | Outer layer | | Modified alloy | | |
| | | Total | Outer layer $d_1$ | Inner layer $d_2$ | Fe | Cr | Fe | Cr | Ni |
| (I) | Oxide-free | / | / | / | / | / | 76 | 17 | 7 |
| (II) | GB pre-oxidized (5 min) | 1.8 | 1.1 | 0.7 | 99 | 1 | 74 | 13 | 12 |
| (III) | GB pre-oxidized (30 min) | 1.7 | 1.0 | 0.7 | 98 | 2 | 75 | 15 | 10 |
| (IV) | GB pre-oxidized + OCP exp. in acid | 1.1 | 0.9 | 0.2 | 43 | 57 | 75 | 15 | 10 |
| (V) | ULP pre-oxidized | 0.9 | 0.4 | 0.5 | 100 | 0 | 81 | 13 | 6 |
| (VI) | ULP pre-oxidized + exp. in air | 1.6 | 0.8 | 0.8 | 100 | 0 | 79 | 13 | 8 |
| (VII) | ULP pre-oxidized + exp. in water | 2.0 | 1.2 | 0.8 | 99 | 1 | 75 | 15 | 10 |
| (VIII) | ULP pre-oxidized + OCP exp. in acid | 1.2 | 1.1 | 0.1 | 39 | 61 | 75 | 15 | 10 |
| (IX) | ULP pre-oxidized + anodic pass. in acid | 2.2 | 0.8 | 1.4 | 60 | 40 | 73 | 15 | 12 |

## Figure 1

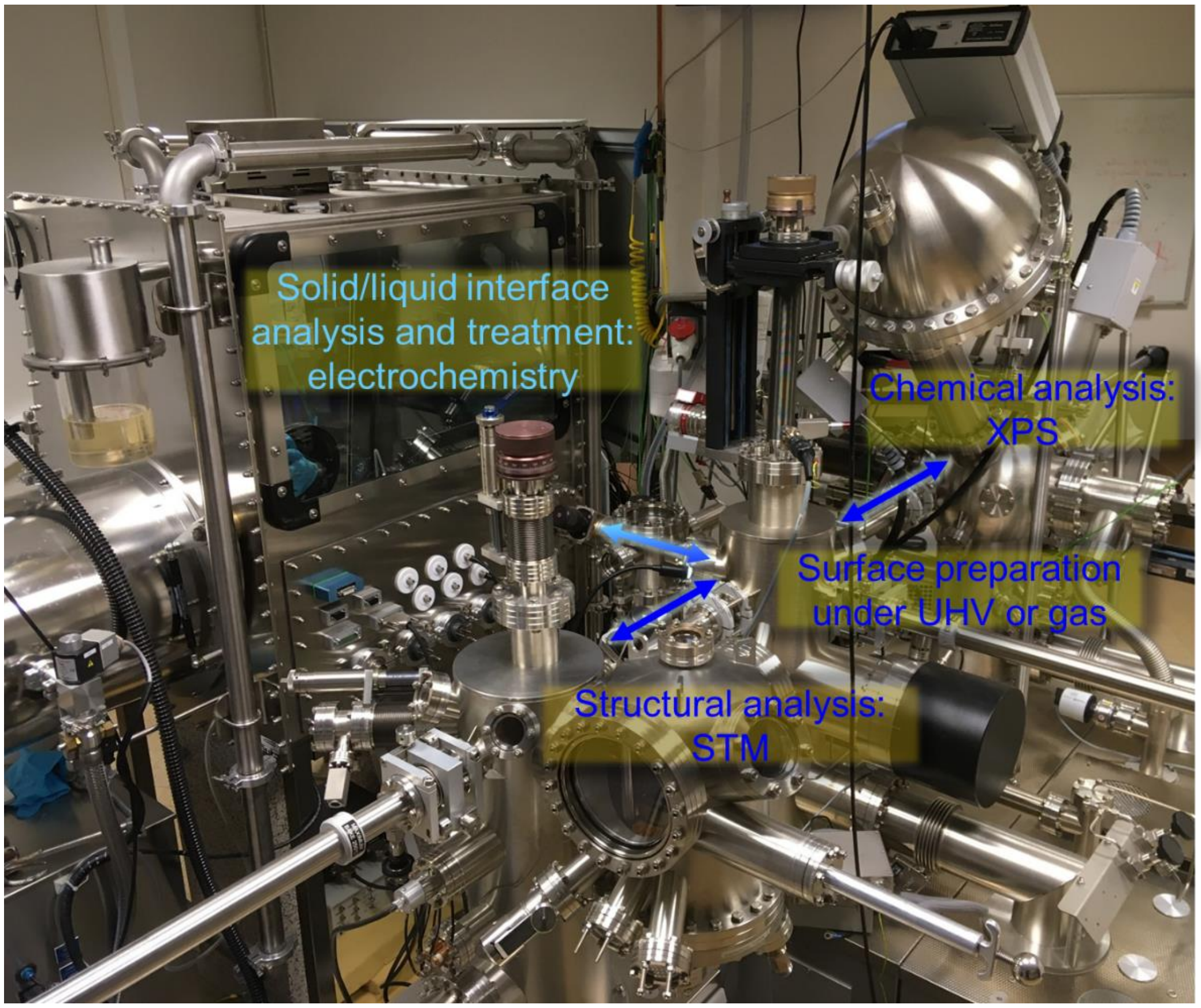
Solid/liquid interface analysis and treatment: electrochemistry
Chemical analysis: XPS
Surface preparation under UHV or gas
Structural analysis: STM

**Figure 2**

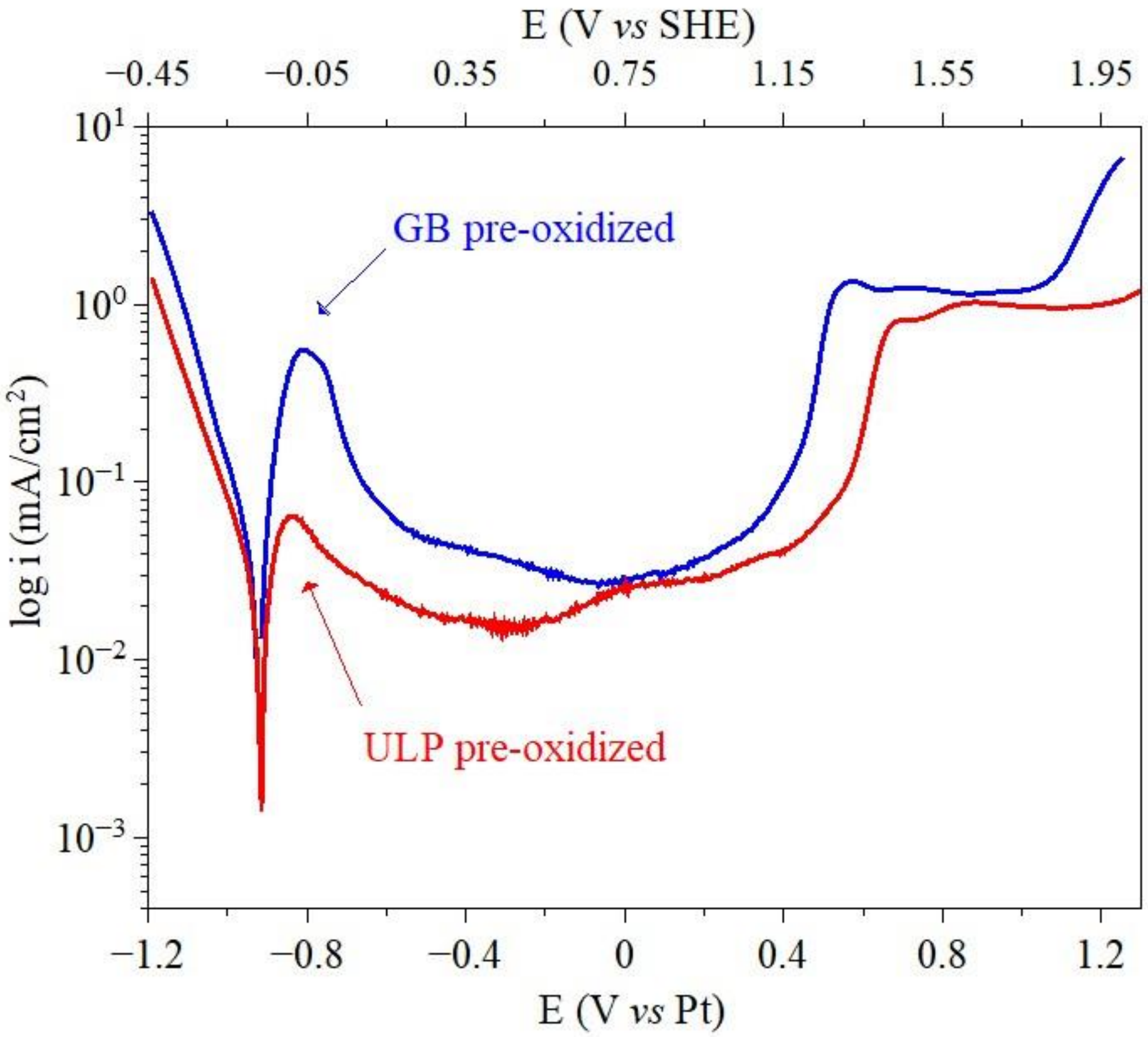

# Figure 3

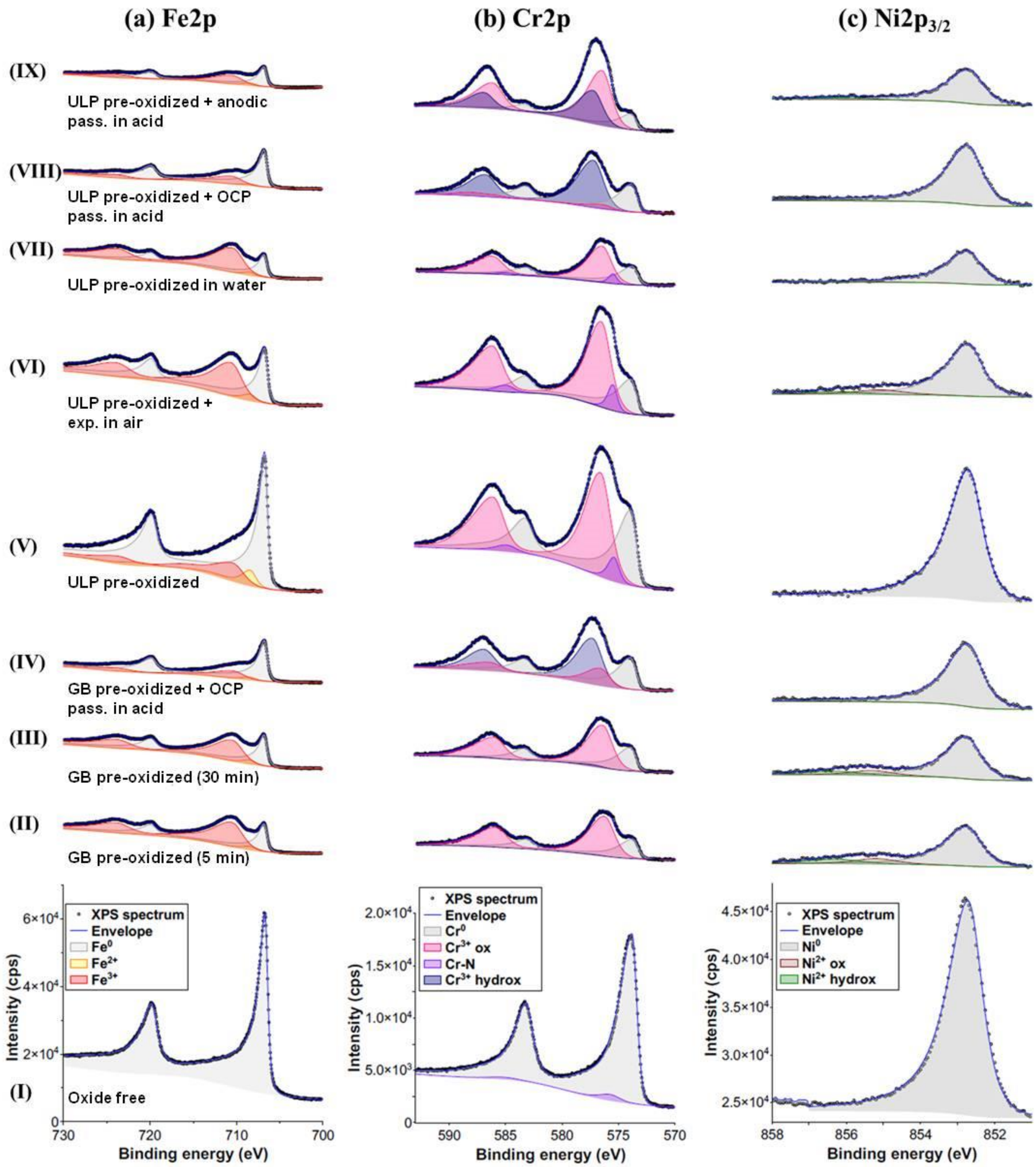

**Figure 4**

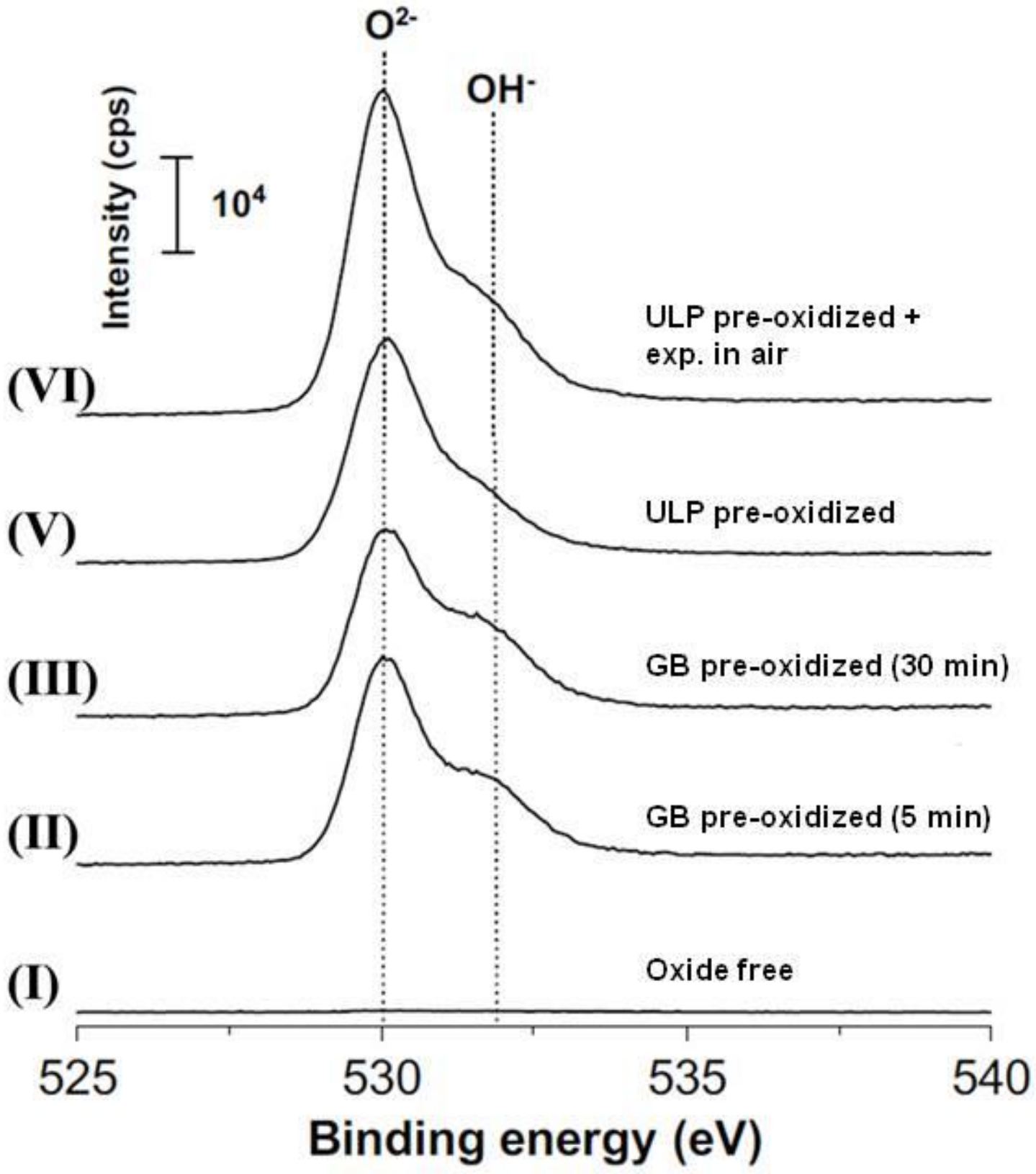

## Figure 5

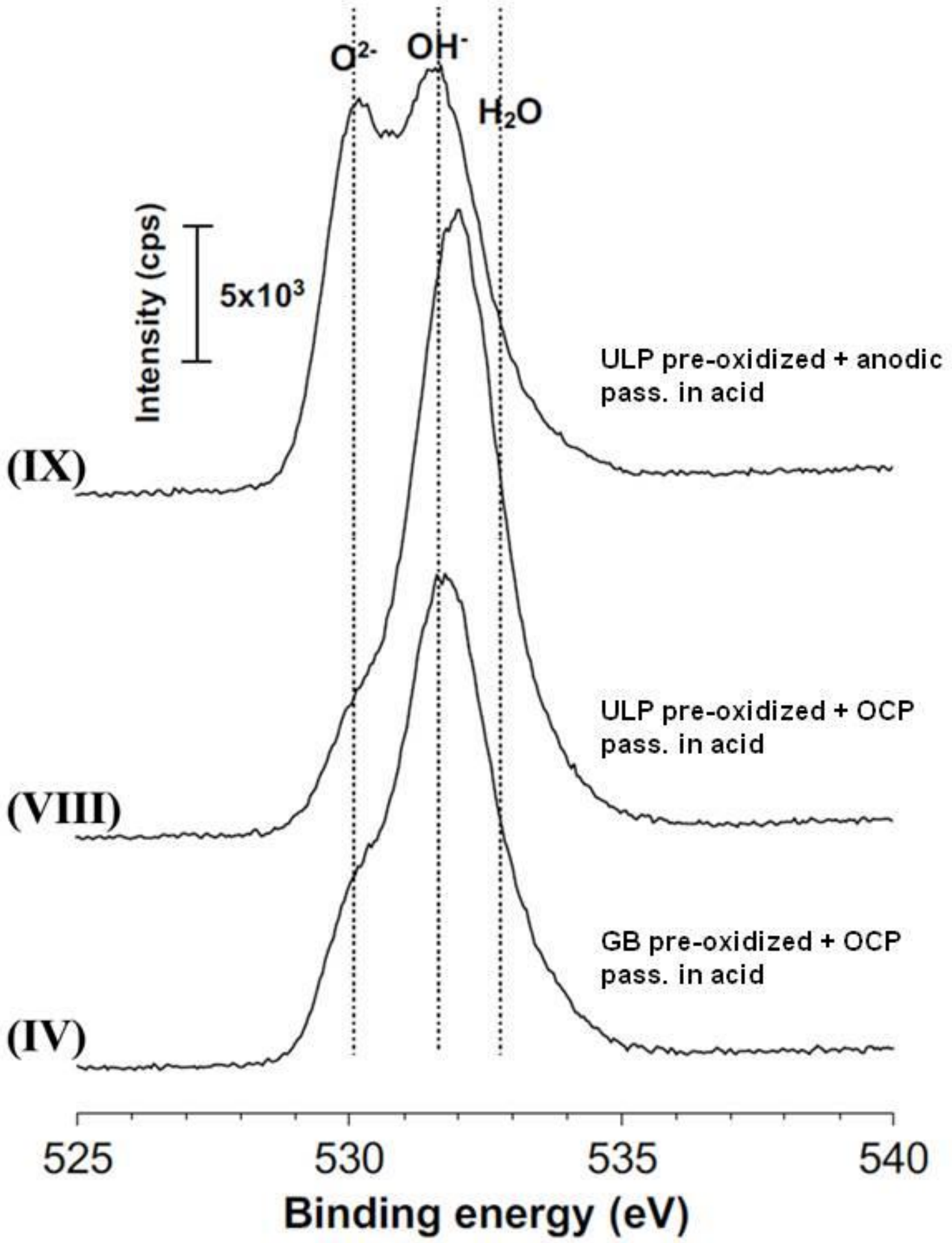

**Figure 6**

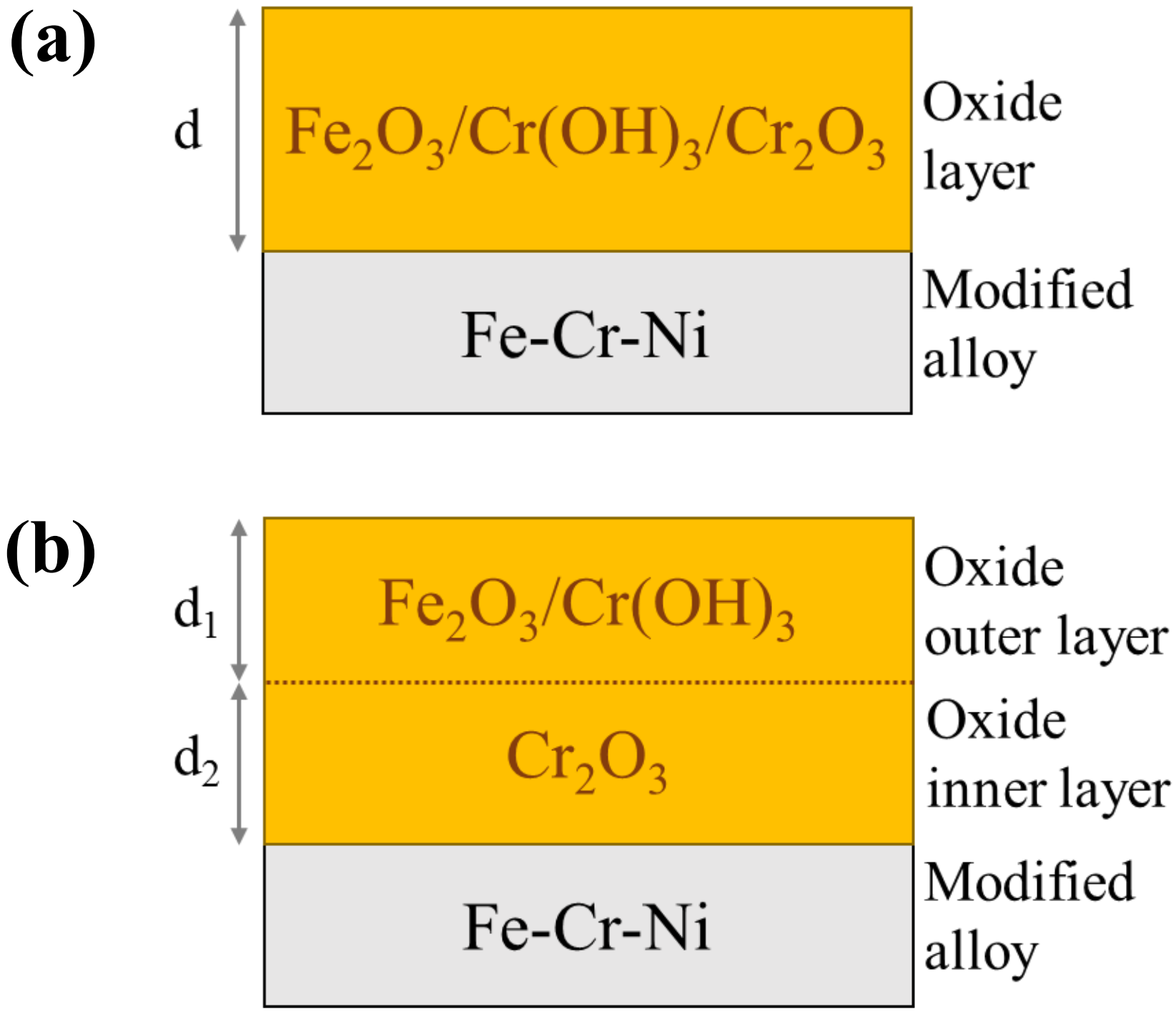

## Figure 7

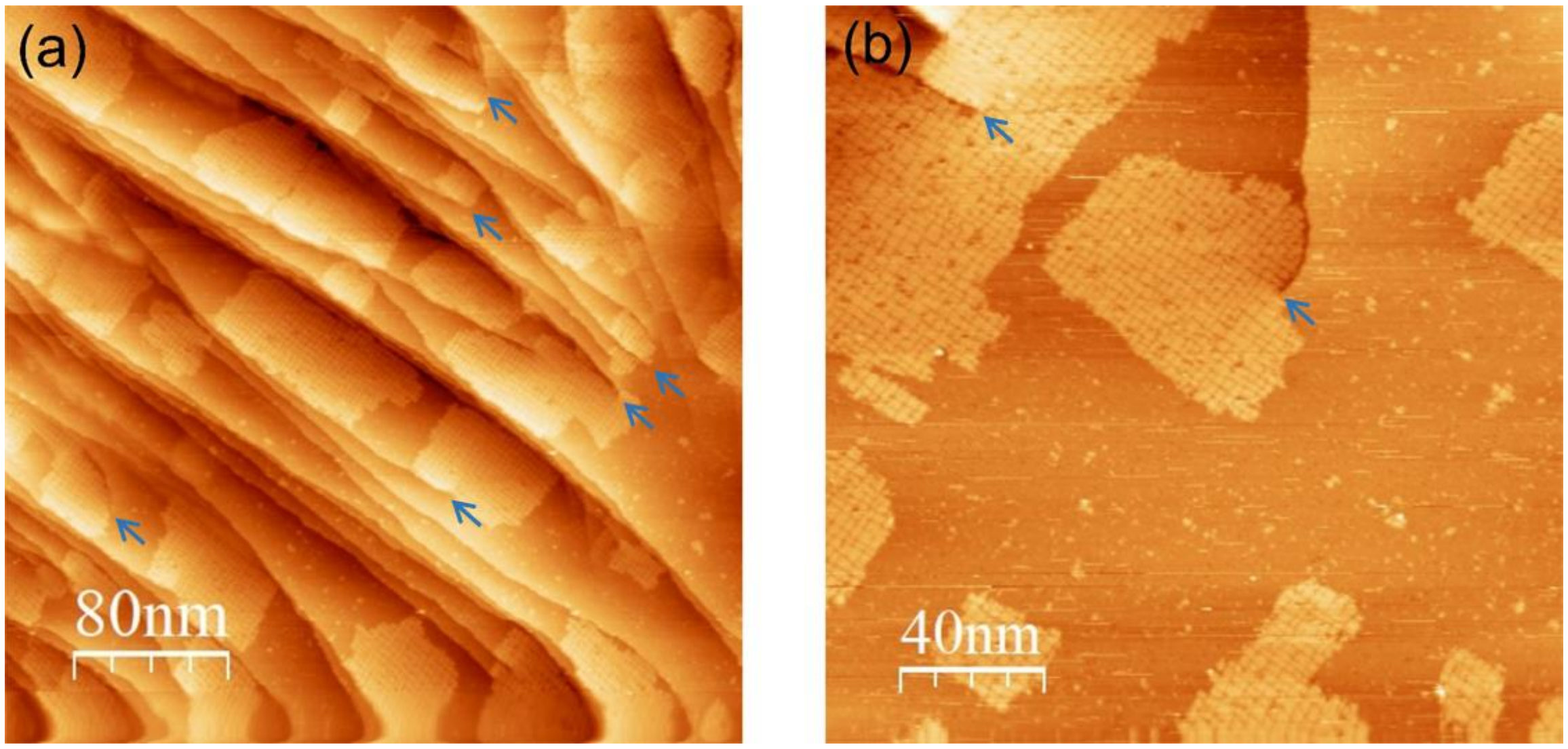

## Figure 8

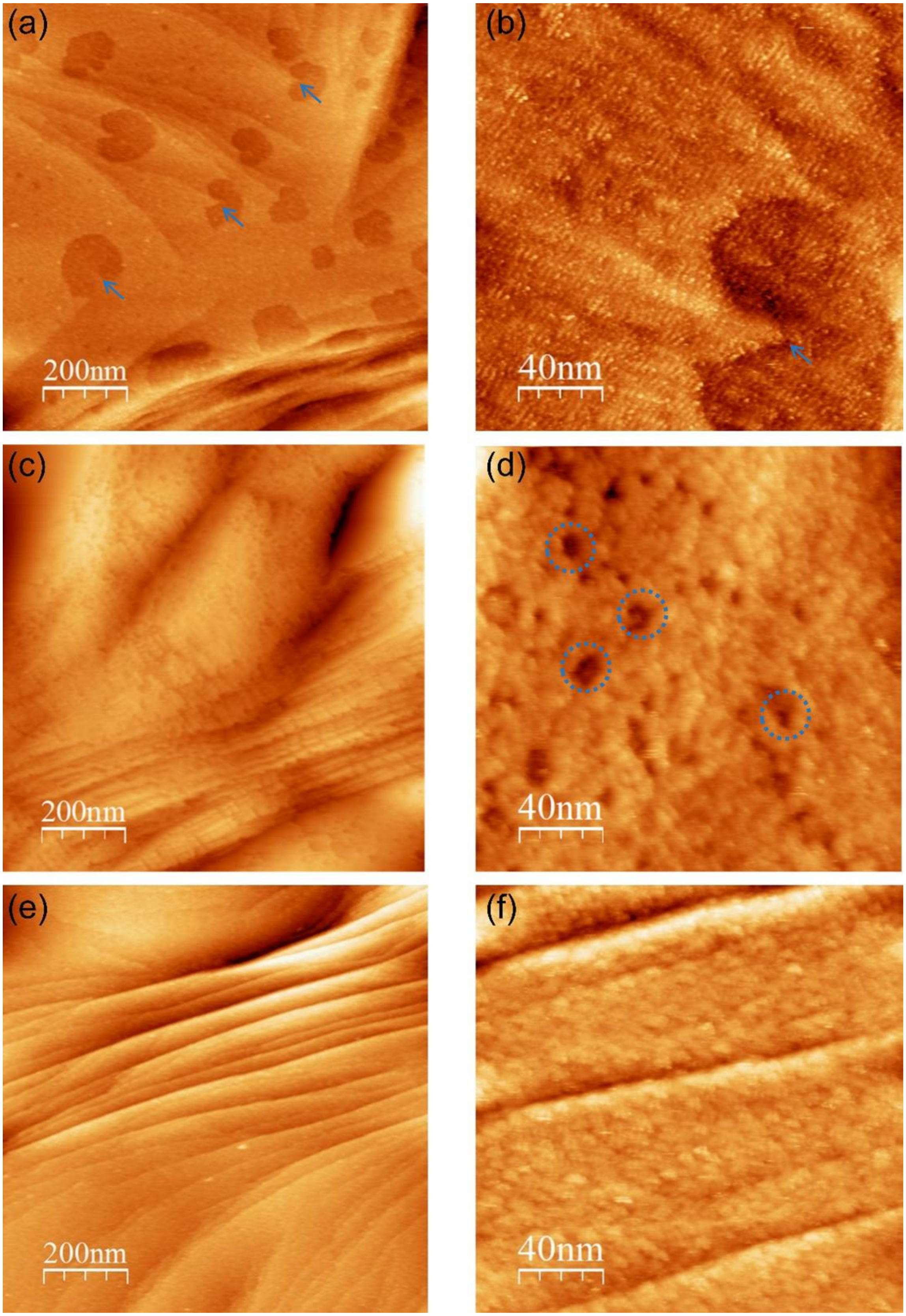

## Figure 9

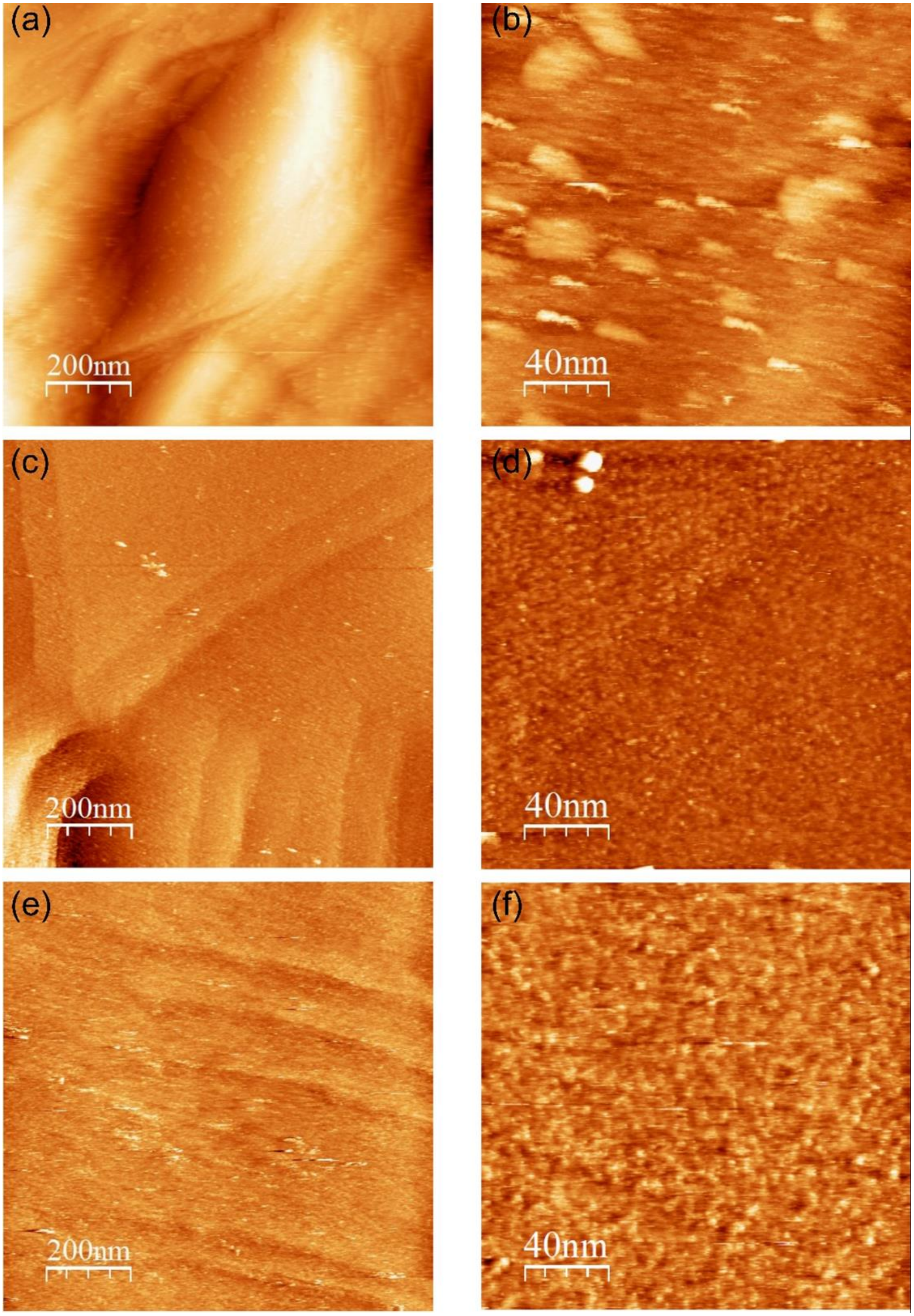

**TOC Figure**

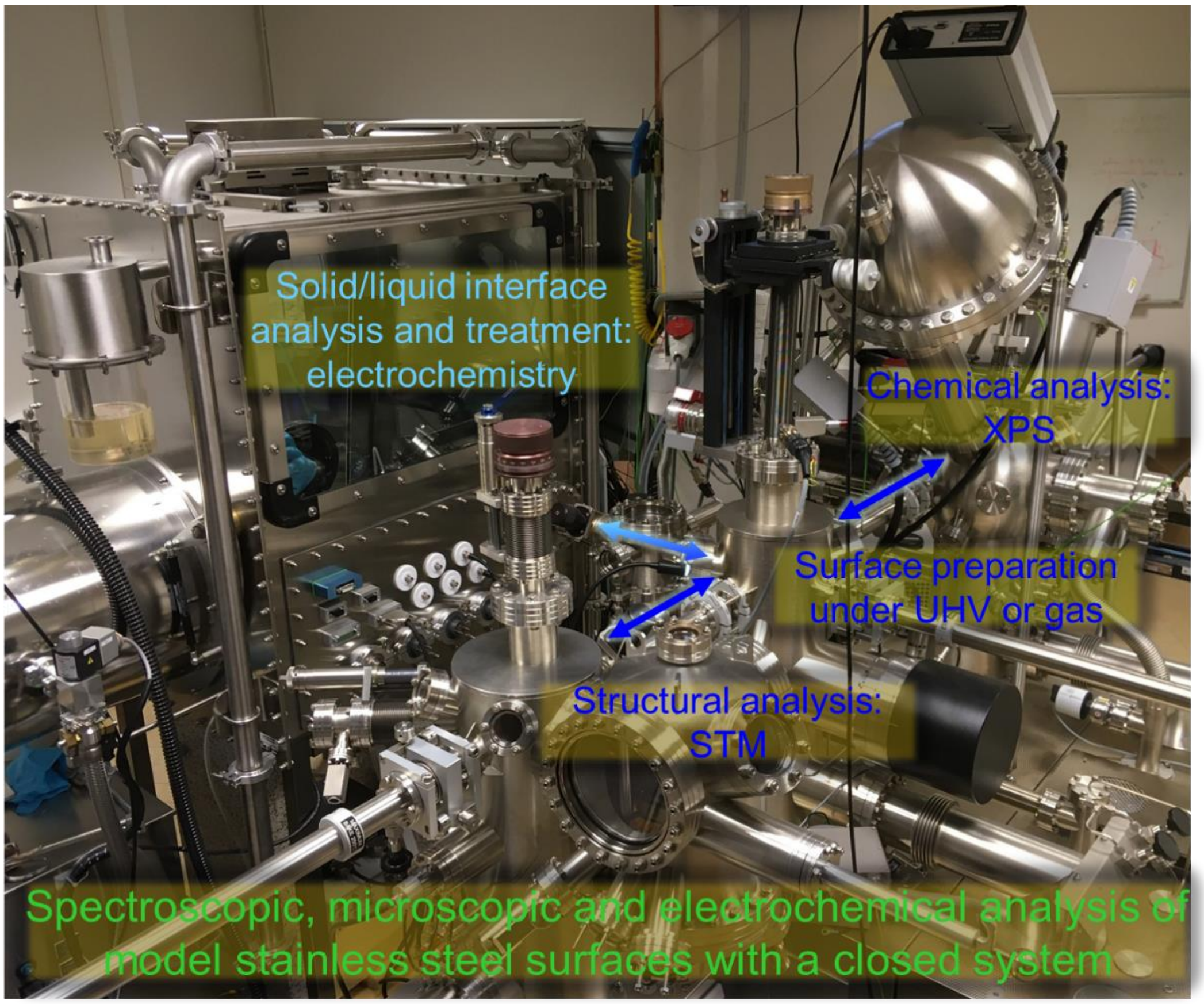
Solid/liquid interface analysis and treatment: electrochemistry
Chemical analysis: XPS
Surface preparation under UHV or gas
Structural analysis: STM
Spectroscopic, microscopic and electrochemical analysis of model stainless steel surfaces with a closed system